\documentclass{ws-procs961x669}            

\usepackage{xspace}	

\newcommand*{\FigPath}{.}
\newcommand*{\BibPath}{.}

\newcommand{\pout}{\mbox{$p_{\rm{out}}$}\xspace}
\newcommand{\pttrig}{\mbox{$p_T^{\rm{trig}}$}\xspace}
\newcommand{\ptassoc}{\mbox{$p_{T}^{\rm assoc}$}\xspace}
\newcommand{\dphi}{\mbox{$\Delta\phi$}\xspace}
\newcommand{\xe}{\mbox{$x_E$}\xspace}
\newcommand{\Ncoll}{\mbox{$N_{\rm coll}$}\xspace}

\begin{document}
\title{Searching for TMD-factorization breaking in $p+p$ and $p$+A collisions:\\
Color interactions in QCD }

\author{C. A. Aidala$^*$}

\address{Physics Department, University of Michigan,\\
Ann Arbor, Michigan 48109, USA\\
$^*$E-mail: caidala@umich.edu}

\begin{abstract}
Dihadron and isolated direct photon-hadron angular correlations have been measured in $p+p$ and $p$+A collisions to investigate possible effects from transverse-momentum-dependent factorization breaking due to color exchange between partons involved in the hard scattering and the proton remnants. The correlations are sensitive to nonperturbative initial-state and final-state transverse momentum $k_T$ and $j_T$ in the azimuthal nearly back-to-back region $\dphi\sim\pi$. In this region, transverse-momentum-dependent evolution can be studied when several different hard scales are measured. To have sensitivity to small transverse momentum scales, nonperturbative momentum widths of \pout, the out-of-plane transverse momentum component perpendicular to the trigger particle, are measured. To quantify the magnitude of any transverse-momentum-dependent factorization breaking effects, calculations will need to be performed for comparison.
\end{abstract}


\bodymatter

\section{Introduction}
One of the frontiers in QCD research is the study of color flow in hadronic interactions. The predicted modified universality of $PT$-odd and $T$-odd correlations in the proton when probed via semi-inclusive deep-inelastic scattering (SIDIS) versus Drell-Yan\cite{Collins:2002kn} is due to different color flow  in these two processes, mediated by gluon exchange between a parton involved in the hard scattering and a proton remnant.  Because of these different color interactions, $PT$-odd and $T$-odd correlations in the proton are predicted to have an opposite sign in the two processes.  The extension of these ideas to hadroproduction of hadrons led to the prediction of TMD-factorization breaking\cite{Bomhof:2006dp,Collins:2007jp,Collins:2007nk,Rogers:2010dm}, with the predicted effect also known as color entanglement.  With gluon exchanges between partons involved in the hard scattering and hadron remnants in both the initial and final state, color flow paths are introduced that cannot be described as flow in the two exchanged gluons separately.  This is a consequence of QCD specifically as a non-Abelian gauge theory, i.e.~due to the fact that gluons themselves carry color.  While breaking of factorization is in fact the rule in QCD interactions, and processes that factorize are the exception, the 2010 prediction by Rogers and Mulders\cite{Rogers:2010dm} is important in that it describes a well-controlled way of breaking factorization, and it in fact goes beyond this, implying novel QCD states of quantum correlated partons across colliding hadrons.  Independent parton distribution functions in the two protons can no longer be used.

In order to search experimentally for the TMD-factorization breaking and color entanglement predicted by Rogers and Mulders, several components are necessary.  
\begin{itemize}
	\item An observable sensitive to a nonperturbative transverse momentum scale as well as a hard interaction scale, such that the TMD-factorization framework would nominally apply.  
	\item Two initial hadrons, such that gluon exchange can occur between a parton involved in the hard scattering and the remnant of the other hadron.
	\item At least one measured final-state hadron, such that gluon exchange can occur between a scattered parton and either remnant.
\end{itemize}

\section{Results}
In proton-proton collisions at PHENIX, such an observable has been developed and measured\cite{Adare:2016bug,Aidala:2018bjf,Aidala:2018eqn}: the out-of-plane momentum component \pout in nearly back-to-back photon-hadron and hadron-hadron production; see Fig.~\ref{fig:kTandpout}a.  The transverse momentum of the "trigger" direct photon or neutral pion, \pttrig, serves as a proxy for the hard scale of the interaction; the out-of-plane momentum component distributions have been measured at several different hard interaction scales.  Figure~\ref{fig:kTandpout}b shows the \pout distributions for both photon-hadron and dihadron correlations, with the shape of the distributions well described by a Gaussian around \pout$=0$, the nonperturbative region.  The distributions transition outward to power-law tails generated by hard (perturbative) gluon radiation.  Note that the curves shown are fits, not phenomenological calculations.  In principle, the measurements can be compared to theoretical calculations in the TMD framework assuming that factorization holds in order to search for TMD-factorization breaking via deviations from the magnitudes, widths, and/or dependence on the hard scale.  The change in the nonperturbative width as a function of the hard scale is of particular interest because the Collins-Soper evolution equation\cite{CS:1981, CS:1982,css_evolution} describing the evolution of TMD distributions with hard scale comes directly from the proof of TMD-factorization\cite{Collins:2012ss}.  While an initial measurement of the Gaussian \pout widths as a function of \pttrig\cite{Adare:2016bug} suggested that they decreased for increasing hard scale, which would be contrary to Collins-Soper-Sterman evolution, the decrease was subsequently understood to be due to the fragmentation kinematics of the away-side particle.  When the fragmentation kinematics are taken into account, the nonperturbative widths instead increase with \pttrig\cite{Aidala:2018bjf}, which is qualitatively similar to what is predicted and observed for Drell-Yan and SIDIS, processes for which TMD-factorization holds.  The Gaussian width of \pout as a function of \pttrig when controlling for the fragmentation kinematics is shown in Fig.~\ref{fig:widthsvspttrigandNcoll}a.  Similar measurements were performed in proton-aluminum and proton-gold collisions\cite{Aidala:2018eqn}, motivated by the idea that the color field of a nuclear remnant might be stronger and lead to larger TMD-factorization breaking effects.  While there is no conclusive evidence for TMD-factorization breaking in these measurements, a broadening of the nonperturbative transverse momentum widths in nuclear collisions was observed, as shown in Fig.~\ref{fig:widthsvspttrigandNcoll}b.

\def\figsubcap#1{\par\noindent\centering\footnotesize(#1)}

\begin{figure}[h]%
	\begin{center}
		\parbox{2.2in}{\includegraphics[width=2.2in]{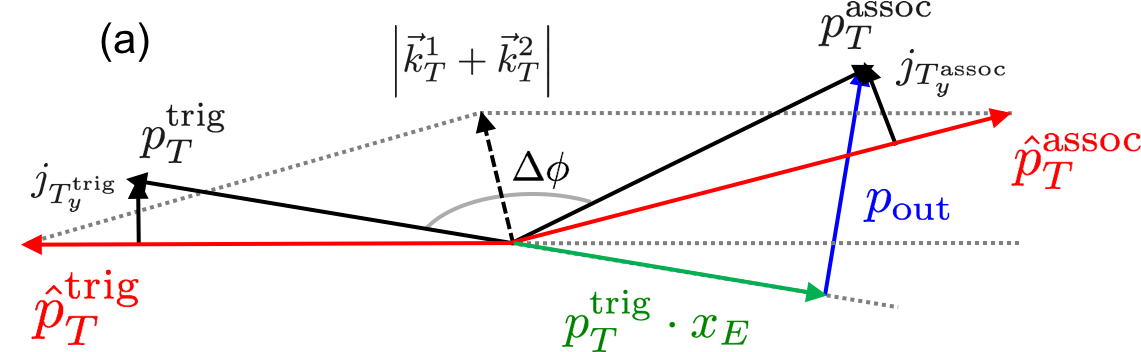}\figsubcap{a}}
		\hspace*{4pt}
		\parbox{2.2in}{\includegraphics[width=2.2in]{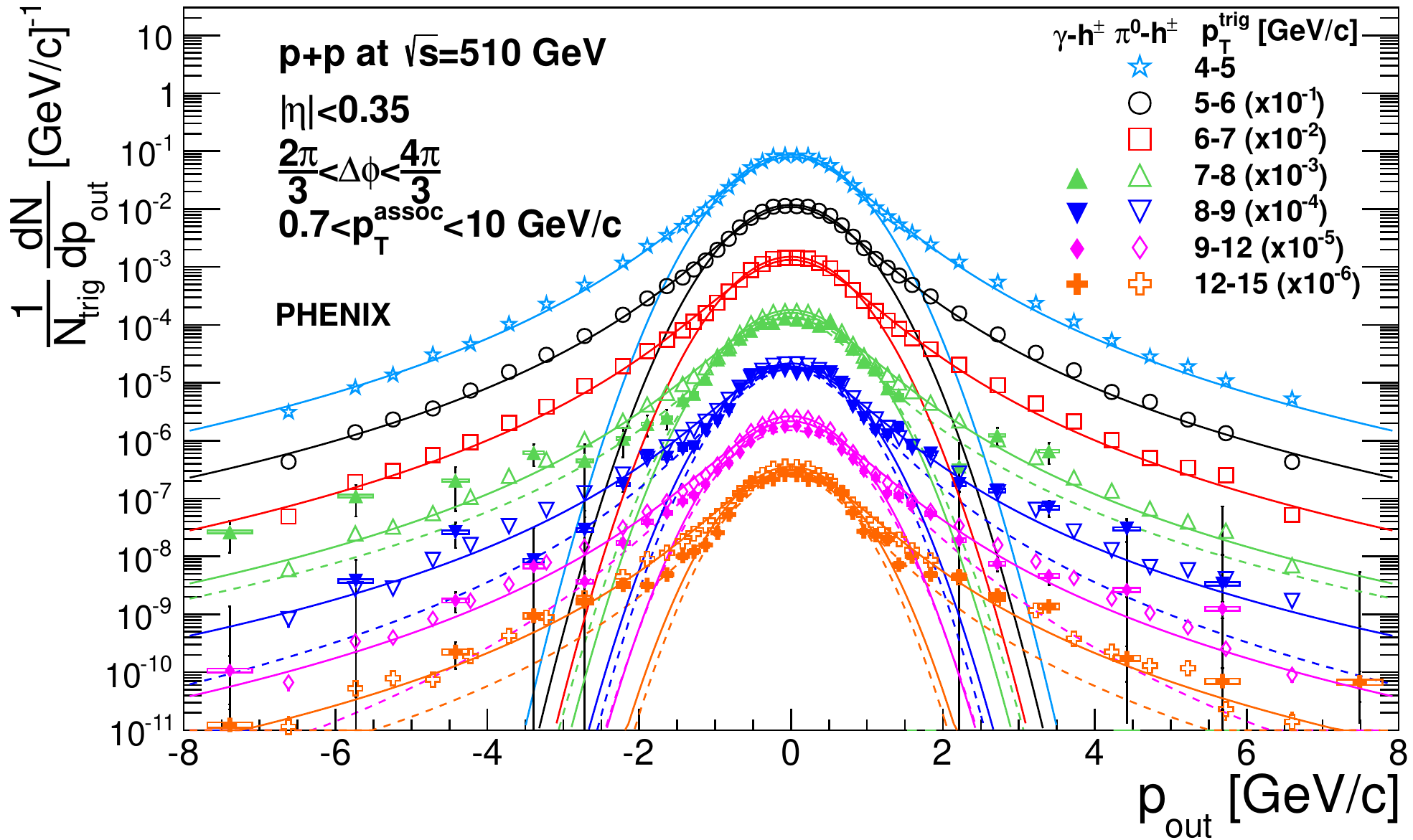}\figsubcap{b}}
		\caption{(a) Hard scattering kinematics of nearly back-to-back dihadrons. Two hard-scattered partons, shown in red, are acoplanar due to the initial-state $\vec{k}_T^1$ and $\vec{k}_T^2$ of the colliding partons. The partons result in a trigger and associated hadrons with \pttrig and \ptassoc. The quantity $x_E$ approximates the momentum fraction $z$ of the final-state away-side hadron\cite{Aidala:2018bjf}.  (b) Per-trigger yields of charged hadrons as a function of \pout. The 
			dihadron and direct photon-hadron distributions are fit with Gaussian functions at 
			small \pout and Kaplan functions over the whole range, showing the 
			transition from nonperturbative behavior generated by initial-state $k_T$ 
			to perturbative behavior generated by hard gluon radiation\cite{Adare:2016bug}.}%
		\label{fig:kTandpout}
	\end{center}
\end{figure}

\def\figsubcap#1{\par\noindent\centering\footnotesize(#1)}
\begin{figure}[h]%
\begin{center}
  \parbox{2.2in}{\includegraphics[width=2.2in]{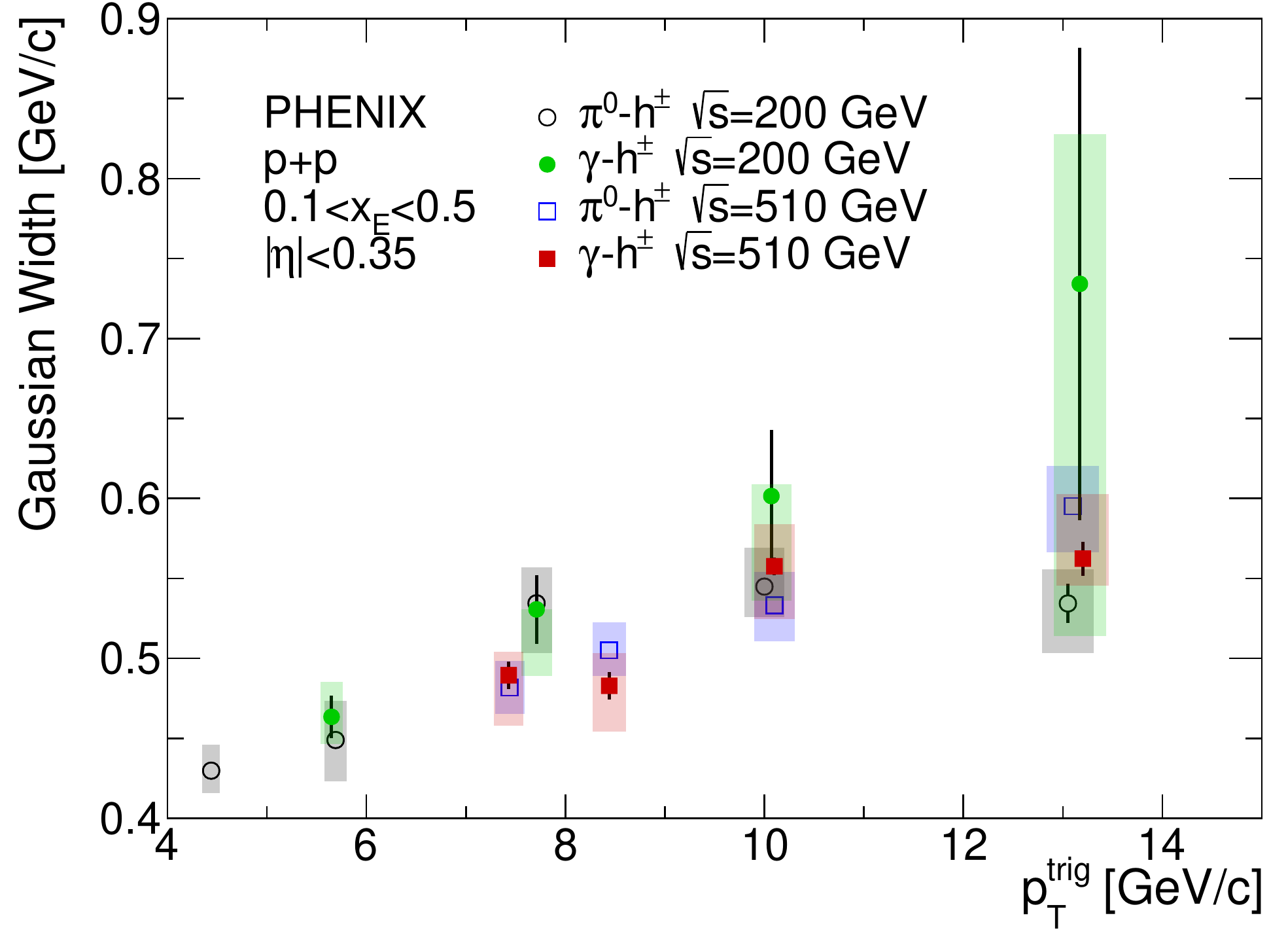}\figsubcap{a}}
  \hspace*{4pt}
  \parbox{2.2in}{\includegraphics[width=2.2in]{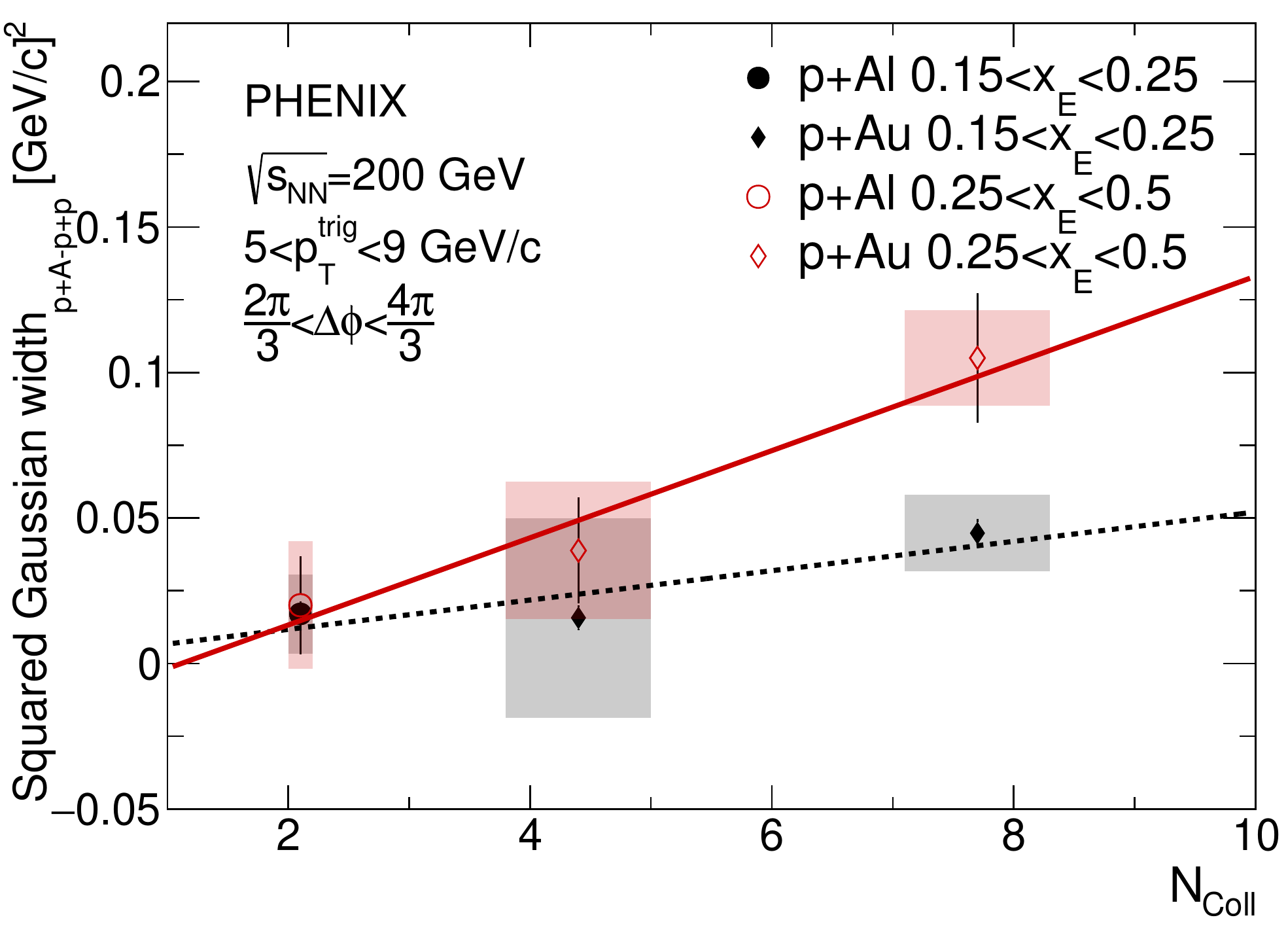}\figsubcap{b}}
  \caption{(a) The Gaussian widths extracted from the \pout distributions in both $\sqrt{s}=$~200 GeV and $\sqrt{s}=$~510 GeV $p+p$ collisions shown as a function of \pttrig\cite{Aidala:2018bjf}. (b) The Gaussian width differences between $p$+A and $p+p$ shown in two \xe bins as a function of the mean number of binary nucleon-nucleon collisions \Ncoll\cite{Aidala:2018eqn}. Linear fits are shown for each \xe bin.}
  \label{fig:widthsvspttrigandNcoll}
\end{center}
\end{figure}

\section{Outlook}
Follow-up measurements are underway in $Z$-hadron, dihadron, and $Z$-jet correlations at the LHCb experiment; $Z$-jet correlations constrain the momentum fractions $x_1$, $x_2$ of the colliding partons and remove sensitivity to any details of hadronization.  Analogous measurements of Drell-Yan and $Z$ dileptons are also planned at LHCb for direct comparison to the $Z$-jet results in overlapping kinematics.  

The community is in the early years of exploring color flow in hadronic scattering processes, and there will remain a great deal to learn from measurements at the future Electron-Ion Collider (EIC).  Color coherence in the form of increased soft hadron production in the region between a high-transverse-momentum hadron or jet and the beam could for example be studied, following earlier measurements in $e^+e^-$ and hadronic collisions (see e.g.~Ref.~\citenum{Chatrchyan:2013fha} and references therein).  In general, color flow is tightly connected to color neutralization and hadronization.  With excellent coverage across the current and target fragmentation regions, at the EIC it may also be possible to examine color connections between these regions by performing more exclusive measurements.   More broadly speaking, studying different hadronization mechanisms, such as vacuum hadronization through string-breaking-type processes, hadronization in a nuclear medium, and recombination of partons moving nearby in phase space in the target region, may provide further insights into color flow and different color-neutralization mechanisms.

\section*{Acknowledgments}
The author acknowledges support from the Office of Nuclear Physics in the Office of Science of the Department of Energy under Grant No.~DE-SC0013393.


\bibliographystyle{ws-procs961x669}
\bibliography{\BibPath/Aidala-INT-18-3}

\end{document}